\documentclass[prb,twocolumn,superscriptaddress,floatfix]{revtex4-1}
\usepackage{graphicx}
\usepackage{amsmath,amssymb}
\usepackage{subfigure}
\usepackage[dvipsnames]{color}
\definecolor{myblue}{named}{MidnightBlue}
\usepackage[a4paper=true,colorlinks=true,citecolor=myblue,linkcolor=black,urlcolor=myblue]{hyperref}
\usepackage{placeins}
\usepackage{bbold}
\usepackage{tabularx}

\newcommand{\be}{\begin{equation}}
\newcommand{\ee}{\end{equation}}
\newcommand{\ba}{\begin{eqnarray}}
\newcommand{\ea}{\end{eqnarray}}

\begin{document}

\title{Tunable single-photon heat conduction in electrical circuits}

\author{P. J. Jones}
\affiliation{Department of Applied Physics/COMP, Aalto University, PO Box 13500, 00076 Aalto, Finland.}
\author{J. A. M. Huhtam\"aki}
\affiliation{Department of Applied Physics/COMP, Aalto University, PO Box 13500, 00076 Aalto, Finland.}
\author{M. Partanen}
\affiliation{Department of Applied Physics/COMP, Aalto University, PO Box 13500, 00076 Aalto, Finland.}
\author{K. Y. Tan}
\affiliation{Department of Applied Physics/COMP, Aalto University, PO Box 13500, 00076 Aalto, Finland.}
\author{M. M\"{o}tt\"{o}nen}
\affiliation{Department of Applied Physics/COMP, Aalto University, PO Box 13500, 00076 Aalto, Finland.}
\affiliation{Low Temperature Laboratory, Aalto University, PO Box
13500, 00076 Aalto, Finland.}

\begin{abstract}
We build on the study of single-photon heat conduction in electronic circuits taking into account the back-action of the superconductor--insulator--normal-metal thermometers. In addition, we show that placing capacitors, resistors, and superconducting quantum interference devices (SQUIDs) into a microwave cavity can severely distort the spatial current profile which, in general, should be accounted for in circuit design. The introduction of SQUIDs also allows for in situ tuning of the photonic power transfer which could be utilized in experiments on superconducting quantum bits.
\end{abstract}
\maketitle

\section{Introduction}
In the past decade great progress has been achieved in the fields of circuit quantum electrodynamics (\onlinecite{Blais04,Wallraff04,Blais07,Deppe08, Fink08,Mariantoni11,Paik11,Niemczyk10}) and photonic heat conduction (\onlinecite{Meschke06,Ojanen07,Cleland04, Pascal11,Muhonen12}). The first attempts to combine them by studying single-photon heat conduction in a microwave cavity were recently reported in Ref.~\onlinecite{CavRevI}. This approach has the potential to deliver practical benefits, both for investigating fundamental quantum phenomena and in order to deliver the practical tools needed in the field of superconducting quantum computing where the manipulation of microwave photons is becoming increasingly important\cite{Houck07,Cleland08,Cleland09,Schoelkopf07,Wallraff11,Chen11}. 

 In Ref.~\onlinecite{CavRevI}, a model to study single photons in a well-defined environment was introduced. The practical setup proposed allows for heat exchange between two normal-metal resistors in a superconducting cavity, for which the photonic heat conduction dominates. This scheme involves a pair of superconductor--insulator--normal-metal (SIN) tunnel junctions \cite{Giazotto06} to control
the temperature of one of the resistors, enabling remote heating or cooling of low-temperature circuit elements. The coupling of the resistors to the superconducting cavity also allows them to act as engineered and controllable environments for the cavity modes.
 In practice, controlling and measuring the temperatures of the resistors inevitably introduces additional power sources for the electron clouds in the resistors. Additionally, there may also be heat transfer between the resistors by quasiparticle excitations in the superconductor. In this paper, we extend the analysis of the system in Ref. \onlinecite{CavRevI} by accounting for the SIN junctions and quasiparticle power sources and show that the observation of single-photon heating and cooling is still experimentally feasible. In addition, this more comprehensive model allows us to impose a cutoff-temperature below which cooling is not observable.

In the second half of this paper, we return to the classical transmission line framework\cite{Pascal11,CavRevI} in order to investigate how placing capacitors, resistors, and superconducting quantum interference devices (SQUIDs) into the transmission line affects the photon mode profile. We show that the distortion of the photon modes responsible for heat transport can be dramatic, and should potentially be accounted for in typical circuit design. Furthermore, this mode distortion suggests a resolution to the discrepancy between the power transfer calculated with the quantum and classical models in the strong coupling limit, which was observed in Ref.~\onlinecite{CavRevI}. Since many future cavity photon experiments\cite{Mariantoni12} will utilize two or more cavities, e.g., for filtering or isolating modes, we also focus on the effect of placing a capacitive break into the central superconducting strip of the transmission line.

We study the possibility of using SQUIDs for in situ tuning of the single-photon power in a cavity. We further develop the classical transmission line model, and demonstrate that such tuning is indeed possible and can be rather effective. In the experiment of Ref.~\onlinecite{Meschke06}, in situ tuning of photonic heat conduction was demonstrated experimentally by using SQUIDs to vary the impedance of the connecting circuit. However, no cavity was present and hence a lumped element model for the impedance was sufficient. In superconducting quantum computing, SQUIDs are often placed into a cavity in order to create or readout quantum bits, qubits\cite{Yang03,Zhou02,Chiorescu04}. Introducing SQUIDs into the line also provides some control over the eigenfrequencies of the photons in the cavity so that they may be, e.g, tuned in and out of resonance with a qubit. This tuning has been demonstrated experimentally through two approaches: either the SQUIDs terminate the transmission line, thus giving variable boundary conditions \cite{Beltrana07,Sandberg08,Wilson11,Bajjani11}, or they are placed in the center conductor of the line\cite{Bertet08}. In Ref.~\onlinecite{Wilson11}, the dynamical Casimir effect was demonstrated; photons were created in a SQUID-terminated transmission line by rapid variation of the SQUID flux. Tunable resonators are also required in several other practical applications: in Ref.~\onlinecite{Bajjani11}, two photon number states are held in a cavity at different frequencies allowing for a parametric interaction between them.

This paper is outlined as follows. In Sec. \ref{sec:power}, we review the basic theory required to calculate the heat power between two normal-metal islands in the center conductor of the line and the theory of the SIN junctions needed to measure their temperature. We also discuss the heat power due to quasiparticles between two such islands into the model. We then study the remote heating and cooling in the presence of SIN junctions in Sec. \ref{sec:repeat1}. Section \ref{sec:capacitor} gives a brief discussion of the discrete transmission line model and the mode profiles are solved with a capacitor placed into the line. In Sec. \ref{sec:resistor}, the model is expanded to include resistors and the power transfer between them is calculated.
In Sec. \ref{sec:SQUID}, we analyze the case of SQUIDs in the line by demonstrating the in-situ tuning of the cavity eigenfrequencies and power transfer.

\section{Comparison of the thermal power sources} \label{sec:power}
\subsection{Phononic and cavity photonic power}
\begin{figure} 
\includegraphics[height=0.4\textwidth]{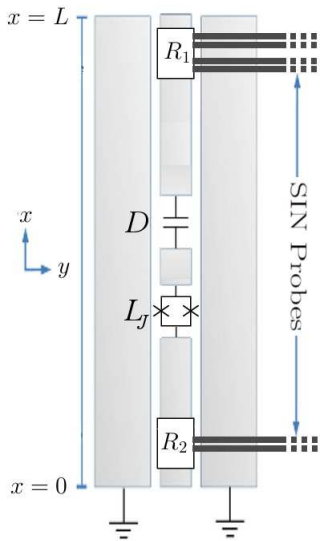}
\caption{Schematic figure of two resistors, $ R_{1}$ and $R_{2}$, a capacitor with capacitance $ D $ and a SQUID with effective inductance $ L_{J} $ in a coplanar waveguide cavity. The SIN probes are also indicated. Both resistors have a pair of thermometer probes and $ R_{1} $ has also a pair of refrigeration probes. This is an extension of the setup to study heat transfer in a cavity proposed in Ref.~\onlinecite{CavRevI}.}
\label{fig:setup}
\end{figure}

In Ref.~\onlinecite{CavRevI}, the heat exchange between two normal-metal resistors in a superconducting cavity was studied in a setup similar to Fig.~\ref{fig:setup} with no SQUID or capacitor in the line, i.e., $ L_{J}=0 $ and $ D\rightarrow \infty $. 
The equilibrium temperature of the second resistor, $ T_{2} $, was calculated when the temperature of the first resistor, $ T_{1} $, was held constant. Power from only two sources was included: the coupling between the electrons in a resistor with the phonons at the bath temperature $ T_{0} $, $ P^{(i)}_{\Sigma}= \Sigma V (T_{i}^{5}-T_{0}^{5}) $, and photon exchange between the resistors and the cavity, $ P_{\Gamma} $. The latter can be derived in the weak coupling limit where the Hamiltonian for a single resistor in a cavity, at position $ x $, may be expressed as
\be  \label{eq:Hint}
\hat{H}(x) = \hat{Q}(x) \otimes \delta \hat{V}_{\textrm{res}},
\ee
where $ \delta \hat{V}_{\textrm{res}} $ is the total voltage fluctuation across the resistor and $ \hat{Q}(x)$ is the integral of the charge density in the cavity from $ 0 $ to $ x $. Since the coupling between the resistor and the cavity is linear, applying Fermi's golden rule with this Hamiltonian yields a photonic heat power between two such resistors of
\be \label{eq:pcav}
P_{\Gamma} = \sum_{k=1}^{\infty} \hbar \omega_{k} \sum_{n=0}^{\infty} \left(\Gamma^{(2),k}_{n+1 \rightarrow n}- \Gamma^{(2),k}_{n \rightarrow n+1}\right)p^{k}_{n},
\ee
where the first sum is over the cavity modes and the second over the photon number for each mode. The probability that the $ k \textrm{th} $ mode contains $ n $ photons is $ p_{n}^{k} $ and the transition rate for photon number in the $ k \textrm{th}$ mode due to the $ i \textrm{th}$ resistor is $\Gamma^{(i),k}_{n \rightarrow n\pm 1}$. For a transmission line of length $ L $, inductance per unit length $ \ell $, and capacitance per unit length $ c $, the frequency of the $ k\textrm{th} $ mode in the cavity is $f_{k}={k}/(2L \sqrt{\ell c}) $, and the transition rates are given by
 \be  \label{eq:TR}
\Gamma^{(i),k}_{n \rightarrow n\pm 1}=\pm \left[\left(n+\frac{1}{2}\pm \frac{1}{2}\right)\gamma^{(i)}\right] f_{B}(\pm \omega_{k},T_{i}),
\ee
where $\gamma^{(i)}={[2R_{i}\sin^{2}(\pi x_{i} /L)]}/({L \ell})$,  
$ x_{i} $ is the position of resistor $ i=1,2 $ and $ R_{i} $ its resistance. The function $ f_{B}(\omega,T)=1/[\exp(\frac{\hbar \omega}{k_{B}T})-1] $ is the Bose--Einstein distribution function.\\

\subsection{SIN refrigeration}
The most practical method of holding the temperature of the first resistor constant over a range of phonon bath temperatures is to remove or supply hot electrons to or from the normal metal by altering the bias of an SIN junction. Using single-electron tunnelling theory, this refrigeration power can be written as\cite{Averin96}
\small
\begin{equation} \label{eq:pref}
P_\textrm{refr}(V)=\frac{1}{e^{2}R_{T}}\int_{-\infty}^{\infty} \! N(E)(E-eV)[f_{N}(E-eV)-f_{S}(E)] \,dE,
\end{equation}
\normalsize
where $ R_{T} $ is the normal state resistance of the tunnel junction, $ f_{j}(E)=1/[\exp(\frac{E}{k_{B}T_{j}})+1] $ is the Fermi--Dirac distribution function in electrode $ j $ and $ N(E)$ is the superconductor density of states (DOS). Ideally 
there are no states in the superconductor energy gap. In practice however,
it is usual to introduce a small phenomenological smearing parameter $ \gamma $ in the DOS in order to account, for example, for the non-ideal superconductor or for the electromagnetic noise\cite{Dynes84,Pekola10}. The DOS is thus calculated as $N(E)=  \left|{\rm Re}\left[(E+i\gamma)/(\sqrt{(E+i\gamma)^{2}-\Delta^{2}})\right]\right|$. It is straightforward to integrate Eq.~(\ref{eq:pref}) numerically
for a given $ T_{N}$ and $ T_{S} $. \\

\subsection{SIN thermometry}
In addition, another pair of SIN junctions are needed on each resistor in order to measure the temperature\cite{Giazotto06}. If a small constant current bias is maintained across the junction, a measurement of the voltage allows for the temperature of the normal metal to be inferred. The SIN thermometry works on the same physical principle as the refrigeration and hence the current through the junction is\cite{Pekola04}
\begin{equation} \label{Eq:SEI}
I=\frac{1}{eR_{T}}\int_{-\infty}^{\infty} \! N(E)[f_{N}(E-eV)-f_{S}(E)] \,dE.
\end{equation}
This allows the current bias $ I_{B} $ to be converted to an equivalent voltage bias. We assume that the supercondcucting probes are thermalized with the bath at temperature $T_{0}$, and hence, the power introduced by the SIN thermometer can be calculated using Eq.~(\ref{eq:pref}). In practice this thermalization may be achieved using quasiparticle traps.\\

\subsection{Photonic power transfer through the junction leads}
In addition to the heat conduction mechanisms discussed above the refrigeration and thermometer probes also provide a supplementary channel through which unwanted classical photonic heat conduction is possible. This excess photonic heat can occur between the resistors in the cavity or between one resistor and another hot element located elsewhere in the setup. We assume that the sample filtering and shielding is sufficiently good such that the latter can be disregarded and we estimate the former using a circuit model that consists of two resistors connected in series with a capacitor\footnote{This circuit model is a rather crude approximation, nevertheless, it should give a reasonable order of magnitude estimate. In any case, such a model should only overestimate this photonic power contribution}. If the temperatures of the resistors are close to each other, $T_{1}\approx T_{2}$, the power through such a circuit is approximately\cite{Timofeev09}  

\begin{equation}
P_{\textrm{Leads}}=\frac{\pi^{3}k_{B}^{2}}{30\hbar}\left[T_{2}^{2}-T_{1}^{2}\right]\left[\frac{k_{B}RC\left(T_{1}+T_{2}\right)}{2\hbar}\right]^{2},
\end{equation}
where $C$ is the capacitance of the series capacitor which is calculated assuming that each junction has a capacitance of $5\textrm{ fF}$. This power can be reduced by fabricating smaller-area junctions with smaller capacitance.\\

\subsection{Quasiparticle heat transport}
The exponential decrease of quasiparticle population with temperature inside a superconductor implies that their effects should be negligible in this low-temperature regime\cite{Saira12}, but in principle they could make a contribution to the power transfer between the resistors. The heat power between the $i^{th}$ resistor and the quasiparticles is given by\cite{Timofeev09}
\begin{equation}
 P_{QP}^{(i)}(x_{i})=-\kappa_{s}(T)A\frac{dT(x)}{dx}\bigg|_{x=x_{i}},
 \end{equation} 
where $ A $ is the cross-sectional area of the superconducting line and the superconductor heat conductivity is denoted by $ \kappa_{s} $. The temperature profile is governed by the differential equation
\begin{equation}
\frac{d}{dx}\left\{-\kappa_{s}[T(x)]\frac{dT(x)}{dx}\right\}=\alpha(T_{0})\Sigma_{\textrm{Al}}\left[T_{0}^{5}-T^{5}(x)\right],
\end{equation}
where $ \alpha(T_{0}) \approx 0.98\exp\left(-\frac{\Delta}{k_{B}T_{0}}\right)$ is the suppression factor of the electron-phonon coupling and $ \Sigma_{\textrm{Al}} $ a material constant for aluminium. 
\\ 

\subsection{Parameters and results} \label{sec:param}
\begin{figure}[!ht] 
\centering
\includegraphics[width=0.5\textwidth]{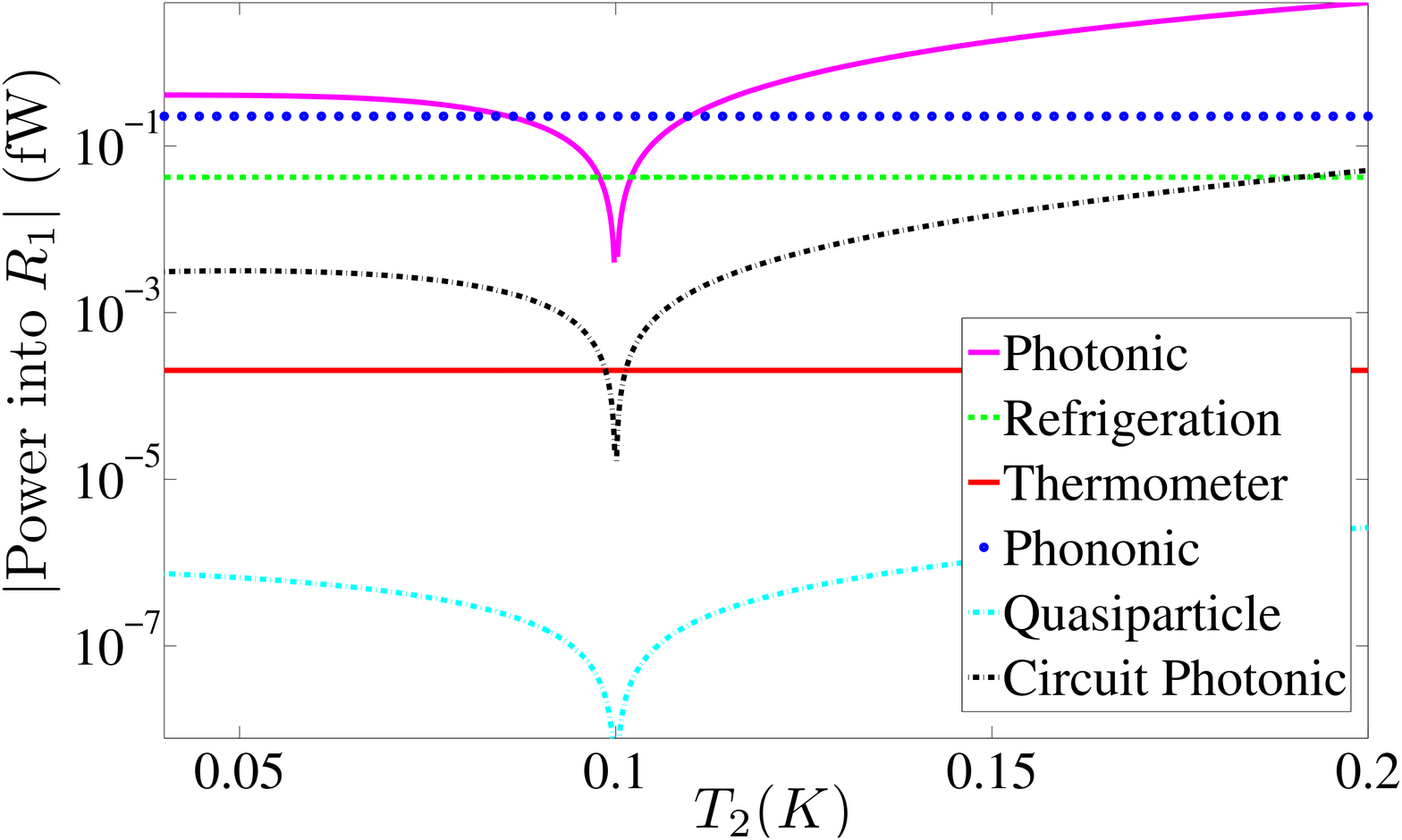}
\caption{The magnitude of the power into Resistor 1 due to the photonic modes (upper solid line), electron--phonon coupling (dotted line), the thermometer (solid line), refrigeration probes (dashed line), junction leads (top dash-dot line) and the excited quasiparticles (bottom dash-dot line). The temperature of Resistor 2 is scanned from $ 40\textrm{ mK} $ to $ 200\textrm{ mK} $. The temperature of Resistor 1 is held at $ 100\textrm{ mK} $ and the bath temperature is $ 80\textrm{ mK} $. The resistors are offset from the ends of the cavity at $ 0.1 \times L $ and $ 0.9 \times L $. The voltage bias across the SIN refrigerator is $ V_{b} =0.6\frac{\Delta}{e}$. All four probes are included. The parameters employed for the cavity are summarized in Table \ref{tbl:parameters}.} \label{fig:pcomp}
\end{figure}

In Fig.~\ref{fig:pcomp}, the power into Resistor 1 from each of the thermal power sources in Sec.~\ref{sec:power} is compared in the case of $ T_{1} = 100\textrm{ mK} $ and $ T_{0}=80\textrm{ mK} $. The temperature of Resistor 2 is scanned from $ 40\textrm{ mK} $ to $ 200\textrm{ mK} $ and the refrigeration bias across a single SIN junction at Resistor 1 is set to $V_{b}=0.6\textrm{ } \Delta/e$. The $230 \textrm{ }\Omega $ resistors have a volume of $ 1.125 \times 10^{-20} \textrm{ m}^{3} $ and are placed in a cavity of length $L= 6.4\times 10^{-3} \textrm{ m} $, where they are offset at $ 0.1 \times L \textrm{ and } 0.9 \times L $ from its ends. For the resistors, we employ the parameters  of $\textrm{Au}_{0.25}\textrm{Pa}_{0.75} $ which we take to have a material parameter of $ \Sigma_{\textrm{Au-Pd}} = 3 \times 10^{9} \textrm{ W} \textrm{m}^{-3}\textrm{K}^{-5}$. The cavity has inductance per unit length of $\ell=4.7\times 10^{-7}\textrm{ Hm}^{-1}$ and a capacitance per unit length of $c=1.3\times 10^{-10} \textrm{ Fm}^{-1}$ with the fundamental mode frequency, $ f_{1}=1 \times 10^{10} \textrm{ s}^{-1} $. The normal-state resistance of the SIN tunnel junctions are $ R_{T} = 19\textrm{ k} \Omega$, the  suppression of the electron--phonon coupling $ \alpha(T_{0}) = 9.1\times 10^{-5} $, and the superconducting heat conductivity, $ \kappa_{s} = 1.7\times 10^{-10}\textrm{ }T(x) \textrm{ W} \textrm{m}^{-1}\textrm{K}^{-2}$ as used in Ref.~\onlinecite{Timofeev09}. We use a thermometer bias current $ I_{b} $ of $ 10\textrm{ pA} $ and take a smearing parameter $ \gamma = 10^{-7} $. For the aluminium center conductor we take $ \Sigma_{\textrm{Al}}=0.3 \times 10^{9} \textrm{ W} \textrm{m}^{-3}\textrm{K}^{-5}$ and an energy gap, $\Delta $ of $ 200\textrm{ } \mu \textrm{eV} $. In this setup, the normal-state resistance of the superconducting line with a cross sectional area of $A=250\times 25\textrm{ nm}^{2}$ is $ R_{l} = 24 \textrm{ k}\Omega $. It is apparent that with these parameters it is the phonon and photon powers that are significant, with the refrigeration and junction lead power being low. We may also conclude that the thermometer power and quasiparticle powers are negligible. However when the magnitude of the bias $|V_{b}|$ is greater than $ \Delta/e $,  i.e., the Fermi level of the resistor is shifted above the superconductor gap, there is a large increase in $ P_{\textrm{refr}} $ and the refrigeration power can be the dominant term. Since Resistor 2 has no refrigeration probe, this only affects Resistor 1. 
\begin{table}[!ht] 
\caption{Parameters of the cavity, materials, and junctions which are assumed in all figures. These parameters are discussed in more detail in Sec.~\ref{sec:param}.  Note that we always use identical resistors.}\label{tbl:parameters}
 \centering  
\begin{tabularx}{0.5\textwidth}{|X|l|}
 \hline
  Quantity & Value \\
  \hline 
  Cavity length, L & $ 6.4 \times 10^{-3} \textrm{ m}$ \\
  
  	Inductance per unit length, $\ell$ & $4.7\times 10^{-7}\textrm{ Hm}^{-1}$ \\
  
   Capacitance per unit length, $c$ & $1.3\times 10^{-10} \textrm{ Fm}^{-1}$ \\
   
  Thermometer Bias Current, $ I_{b} $ & $ 1\times 10^{-11}\textrm{ A} $   \\
  
  Junction lead series capacitance, $C$ & $6.67 \textrm{ fF}$ \\
  
  Cross-sectional area of line, A  & $ 250\times25\textrm{ nm}^{2} $  \\
  
 Smearing parameter, $ \gamma $  & $ 1\times 10^{-7} $   \\
 
 	Resitor volume, $V$ &  $ 1.125 \times 10^{-20} \textrm{ m}^{3} $ \\
 	
 Superconductor conductivity, $ \kappa_{s}(T) $  &  $ 1.7\times 10^{-10} T\textrm{ W} \textrm{m}^{-1}\textrm{K}^{-2}$  \\
  
   Superconducting gap of Al, $ \Delta $ & $ 200\textrm{ } \mu \textrm{eV} $   \\
       
  Al material parameter, $ \Sigma_{\textrm{Al}} $   & $ 0.3 \times 10^{9} \textrm{ W} \textrm{m}^{-3}\textrm{K}^{-5}$  \\  
  
  $\textrm{Au}_{0.25}\textrm{Pa}_{0.75} $ parameter,  $ \Sigma_{\textrm{Au-Pd}} $   & $ 3 \times 10^{9} \textrm{ W} \textrm{m}^{-3}\textrm{K}^{-5}$  \\  
  
      Normal state resistance of line, $ R_{l} $  &  $ 24 \textrm{ k}\Omega $.   \\
  
   Normal state NIS junction resistance, $ R_{T} $  &  $ 19\textrm{ k} \Omega$  \\

  \hline 
\end{tabularx}
   
   \end{table} 
   
\section{Temperature Response of the resistor} \label{sec:repeat1}

\begin{figure}[h!t] 
\centering
\includegraphics[width=0.5\textwidth]{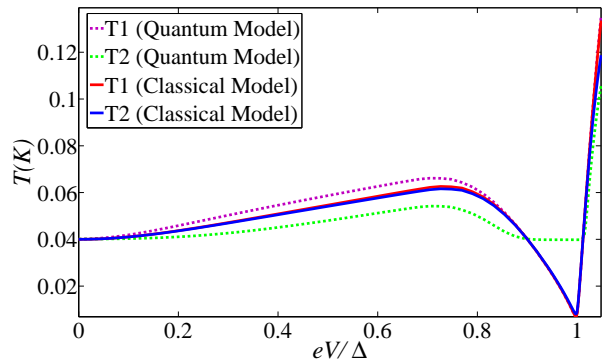}
\caption{Resistor temperatures as functions of the bias voltage on Resistor 1 (solid lines). Results calculated with the classical photonic power in an impedance matched loop used in Ref. \onlinecite{Timofeev09} are also shown (dashed lines). With bath temperature $ T_{0}=40\textrm{ mK} $, deviations in the calculated values of $ T_{2} $, the temperature of the second resistor, are observed between the classical loop and quantum cavity models. The temperature is a symmetric functions of the bias. The parameters employed for the cavity are summarized in Table \ref{tbl:parameters}. }\label{fig:mid}
\end{figure}

In the steady state, there is no net power flow at Resistors 1 and 2, and hence we have the coupled pair of equations
\begin{align} 
2P^{(1)}_{\textrm{refr}}+2P^{(1)}_{\textrm{Therm}}+P^{(1)}_{QP}-P_{\textrm{Leads}}-P_{\Gamma}-P^{(1)}_{\Sigma}=0 
\label{eq:res1} 
\\
2P^{(2)}_{\textrm{Therm}}+P^{(2)}_{QP}+P_{\textrm{Leads}}+P_{\Gamma}-P^{(2)}_{\Sigma}=0 \label{eq:res2}
\end{align}
The factors of 2 arise because there are two thermometer and two refrigerator probes.
We do not impose a priori the restriction that $ T_{1} $ is fixed, but solve Eqs.~(\ref{eq:res1}) and (\ref{eq:res2}) together to find the equilibrium values for $ T_{1} $ and $ T_{2}$ as a function of the refrigerator bias, $ {V}_{b}$, on the first resistor. This is precisely what is done in Ref.~\onlinecite{Timofeev09} using a classical photonic power\cite{Cleland04} for the case of two resistors in an impedance matched superconducting loop. The latter setup allows the full quantum of thermal conductance to be achieved, i.e., the maximum possible power transfer via a single one-dimensional photonic channel. In  Fig.~\ref{fig:mid}, we observe that the temperatures resulting from the classical photon power used in Ref. \onlinecite{Timofeev09} give different results to the quantum photonic power given by Eq.~(\ref{eq:pcav}) at $ T_{0}=40 \textrm{ mK}$.  The heating below the gap is due to the leakage current caused by the smearing of the density of states, and dissapears for $ \gamma \to 0$. \\
Let us simulate the single-photon heat conduction experiment proposed in Ref. \onlinecite{CavRevI} including also the refrigeration, thermometer, junction lead and quasiparticle power sources in addition to the phonon and photonic heat power. Since Eq.~(\ref{eq:res2}) does not depend on the refrigeration bias, with $ T_{1}$ held fixed, it may be solved for $ T_{2} $ at a given $ T_{1} $. With this $ T_{2} $, Eq.~(\ref{eq:res1}) can then be solved to find the required SIN refrigeration voltage to make the equations consistent.
 \begin{figure}[h!t] 
\centering
\includegraphics[width=0.45\textwidth]{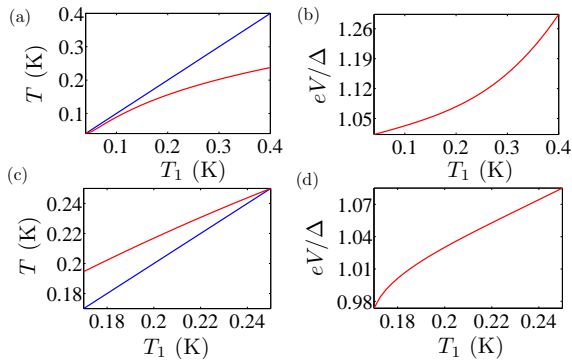}
\caption{ (a) Single-photon heating of Resistor 2 is observed when the temperature of Resistor 1 is swept from $ T_{1}=40\textrm{ mK} $ to $  T_{1}=400\textrm{ mK} $ for the bath temperature $T_{0}= 40\textrm{ mK} $.\\(b) Corresponding heating bias which must be applied to Resistor 1. (c) Single-photon cooling of Resistor 2 is observed when the temperature of Resistor 1 is swept from $ T_{1}=250\textrm{ mK} $ to $  T_{1}=170\textrm{ mK} $ for the bath temperature $T_{0}= 250\textrm{ mK} $. (d) Corresponding cooling bias which must be applied to Resistor 1. Deviations in $T_{2}$ due to the effect of the additional thermometer, junction and quasiparticle power sources are indistinguishable in practice. The parameters employed for the cavity are summarized in Table \ref{tbl:parameters}.}
\label{fig:heating}
\end{figure}
The results are shown in Fig.~\ref{fig:heating}. The inclusion of the additional power sources has very little effect on the equilibrium value of $ T_{2} $. Instead the major consequence of incorporating the probes into the model is that it restricts the regime of consistent solutions for photonic cooling of Resistor 2. With a bath temperature of $ 250\textrm{ mK} $ it is possible to find solutions only for $ T_{1} \gtrsim 170\textrm{ mK} $. Nevertheless, the temperature range is sufficient 
for the observation of single-photon cooling, as shown in Fig.~\ref{fig:heating}(c). The bias voltages $ {V}_{b}$ needed to make the equations consistent are comfortably achievable in practice\cite{Timofeev09}.

\section{Photon Mode Distortion Due To a Capacitor} \label{sec:capacitor}
\begin{figure}[h!]
\includegraphics[width=0.5\textwidth]{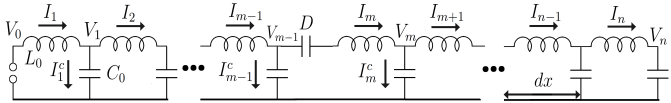}
\caption{Circuit model of a transmission line cavity with a capacitor in the center conductor at position $ m $. The internal resistance of the superconducting line may be safely neglected in our case. Thus the basic transmission line may be modelled as an infinite sequence of inductors and shunt capacitors. We introduce a capacitor of size $ D $ at the position labelled by the index $ m $ into the central conductor. We assume that the inductance and capacitance per unit length, $ \ell $ and $ c $, respectively, are constant along the line and hence for each inductor $ L_{0}=\ell dx $ and capacitor $ C_{0}=cdx $.} \label{fig:Line}
\end{figure}

We first demonstrate that placing electrical components in a cavity changes the spatial profile of the photon modes. This has
profound effects, e.g., on the power transfer. 
We begin by simulating the system in which we partition our transmission line into two coupled cavities.
We employ here the model of the transmission line as an infinite series
of lumped LC elements. In order to divide this waveguide into two seperate cavities 
we add a capacitive break of magnitude $ D $ into the center conductor of the coplanar waveguide as shown in Fig.~\ref{fig:Line}. 

With the capacitor placed at position index $ m $, we look for solutions of the form $ I(x,t)=I(x)\cos(\omega_{i} t+\phi_{i}) $, where $ \omega_{i}=2\pi f_{i}$ and $ \phi_{i} $ are the angular frequency and phase of the mode respectively. By applying Kirchoff's laws at each node we obtain the eigensystem
\begin{eqnarray}
&&(-L_{0}C_{0}\omega_{i}^{2}+2){I}_{n}-I_{n-1}-I_{n+1}=0, \textrm{ for $ n \ne m $}, \label{eq:ind}\\
&&\left(-L_{0}C_{0}\omega_{i}^{2}+2+\frac{C_{0}}{D} \right)I_{m}-I_{m+1}-I_{m-1}=0.  \label{eq:cap}
\end{eqnarray}
Therefore, with no resistors in the cavities, we have an eigenvalue equation for the spatial dependence of the current, $ MI(x_{n})=L_{0}C_{0}\omega_{i}^{2}I(x_{n}) $, where the matrix $ M $ is of the form
\begin{equation}
M=M_{0}+\frac{C_{0}}{D}j^{mm}.
\end{equation}
Here $M_{0}$ is the matrix without the presence of the capacitor which has the from of a discretized Laplacian and $ j^{nm} $ denotes the single-entry matrix with the only non-vanishing element equal to unity at $ (n,m) $.
We observe that the important factor is the ratio of the intrinsic capacitance $ C_{0} $ to the coupling capacitance $ D $. In the limit $ D \to \infty $ we return to the matrix for the unmodified cavity, and in the limit $ D \to 0 $, we have two independent cavities. \\ \\

\begin{figure} 
\includegraphics[width=0.5\textwidth]{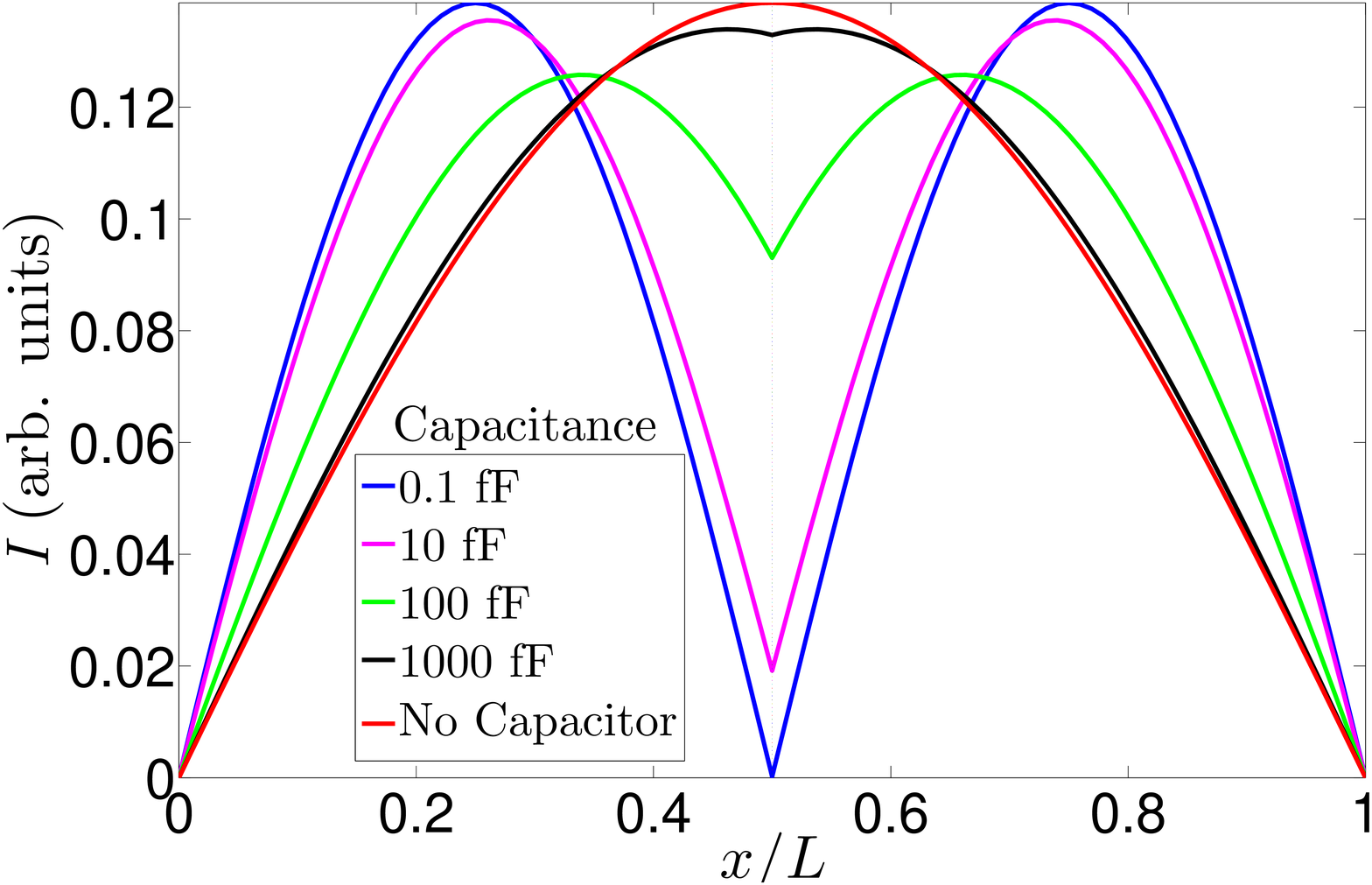}
\caption{Magnitude of the current profile for the fundamental mode with a coupling capacitor of size $0.1, 10, 100 $, and $ 1000\textrm{ fF} $ (from bottom to top at $ x=L/2 $) is placed at $ \frac{L}{2} $. The corresponding values of $ f_{1} $ are $ 1.96, 1.89, 1.45 \textrm{ and } 1.07 \times 10^{10} \textrm{ Hz}  $, respectively. The sinusoidal profile for the unmodified transmission line is also plotted with the eigenfrequency $ f_{1} = 1 \times 10^{10} \textrm{ Hz}  $. The parameters employed for the cavity are summarized in Table \ref{tbl:parameters}.}
\label{fig:1CC}
\end{figure}

This system may be solved analytically for any number of capacitors in arbitrary positions. In the case in which one capacitor is placed at $ L/2 $ we have
\small
\begin{align*}
I(x)=
\begin{cases}
A\sin(kx),\qquad \text{ } \quad  \text{ } \quad  \text{ } \qquad  \text{ for } 0\le x \le L/2 \\
A\sin[k(L-x)], \qquad \text{  } \qquad \text{ }  \text{ for } L/2 < x \le L,
\end{cases}
\end{align*}
\normalsize
where $ A $ is a normalisation constant and the wavevector $ k $ is found from
\be
2k+\frac{c}{D}\tan\left(\frac{kL}{2}\right)=0.
\ee
 Figure~\ref{fig:1CC} shows the effect on the current profile of the fundamental mode when capacitors of various sizes are placed into a transmission line. The spatial profile is a sensitive function of both capacitor size and position. For small capacitance the current modes are vastly altered. As the capacitance is increased the profile tends back towards that of the unmodified line, as expected.  \\

\section{Effect of the capacitor on the power transfer} \label{sec:resistor}
\
Let us add a resistor of size $ R=r\Delta x $ to the circuit of Fig. \ref{fig:Line} at position index $ k $. We repeat the procedure of applying Kirchoff's laws at this node. Due to the finite temperature of the resistor we  associate a fluctuating noise voltage $ \delta V_{k}(t) $
across it. We obtain 
\begin{align}
&&V_{k-1}-\delta V_{k}(t)-I_{k}R-L_{0}\frac{dI_{k}}{dt}-V_{k}=0,\nonumber \\ 
&& \left(L_{0}C_{0}\partial^{2}_{t}+C_{0}R\partial_{t}+2\right)I_{k}+I_{k-1}+I_{k+1} =-C_{0}\delta \dot{V}_{k}(t).  \label{eq:TimeRes}
\end{align}
To study the base effect on the mode profile, we first assume that the resistor does not introduce any noise by setting $\delta V(t)=0 $.
In this case, Eq. (\ref{eq:TimeRes}) along with Eqs. (\ref{eq:ind}) and (\ref{eq:cap}) form the system $ Z(\omega)I(x_{n})=0 $.
Here $ Z(\omega) $ is an impedance matrix given by
\begin{equation} \label{eq:impZ}
Z(\omega)= M_{0}+\frac{C_{0}}{D}j^{mm}- i\omega C_{0}R j^{kk}+LC_{0} \omega^{2},
\end{equation} 
For non-trivial $ I(x_{n}) $ we then have the condition that $ \textrm{det}\left[Z(\omega)\right]=0$, yielding 
the eigenfrequencies of the cavity. This allows $ Z(\omega_{i}) $ to be constructed, from which the current may be found.  \\

\begin{figure} 
\includegraphics[width=0.5\textwidth]{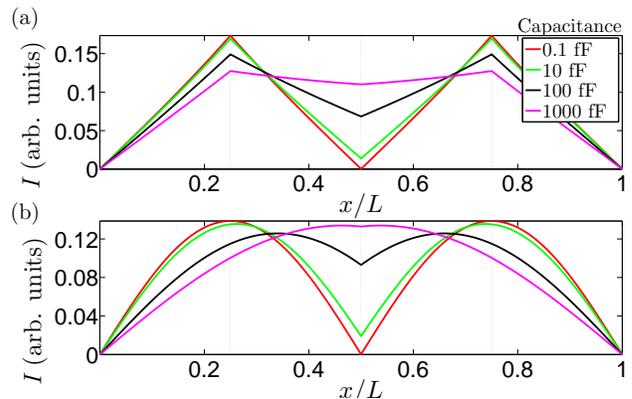}
\caption{(a) Normalised current profile for a coupling capacitor of $0.1, 10, 100 $ and $ 1000\textrm{ fF} $ placed at $ \frac{L}{2} $ and resistors of $ 230 \textrm{ }\Omega $ placed at $ \frac{L}{4} $ and $ \frac{3L}{4} $. Here we assume that the resistor introduces no noise fluctuations into the transmission line. In this case, the coupling is strong enought that the values of $ f_{1} $ are purely imaginary, corresponding to decaying solutions. (b) Normalised current profile for the same system but with $ R= 2.3 \textrm{ }\Omega $. The parameters employed for the cavity are summarized in Table \ref{tbl:parameters}.}
\label{fig:orc}
\end{figure}

Figure~\ref{fig:orc}(a) shows that adding a resistor of $ 230\textrm{ } \Omega $ into the center conductor of the transmission line changes the mode profile substantially. The coupling between the resistor and cavity is strong enough for the eigenfrequencies of the modes to have a dominant imaginary component, hence these solutions are found to be rapidly decaying. In Fig.~\ref{fig:orc}(b), we simulate the same system but with $ R=2.3 \textrm{ }\Omega $. Comparison of the two figures shows that the distortion of the mode is almost negligible in the latter case. This is likely to be the reason for the previously observed discrepancy between the photonic power transfer in the classical and quantum models in the strong-coupling regime\cite{CavRevI}. In order to remain in the weak-coupling limit, either the magnitude of the resistor should be small or the resistor should be positioned in the cavity at a point where the amplitude of the mode is small. In Sec. \ref{sec:repeat1} and Ref. \onlinecite{CavRevI}, the constraint that the resistors must remain close to the ends of the cavity ensures this. Otherwise, even in these simple setups it is possible to create vastly different spatial mode profiles simply by altering the position and magnitude of the components.\\ \\
\noindent
Next we account for the voltage noise from the resistors. To handle $ \delta V(t) $ we move into the frequency domain, where the current is governed by the matrix equation $Z(\omega)I(\omega)=i\omega C_{0}\delta V(\omega) $.
In our case, the fluctuations are due to the Johnson--Nyquist noise, with the power spectrum
\begin{equation}
S_{V}(\omega ,T)=\frac{2 R \hbar \omega}{1-\exp(\frac{-\hbar \omega}{k_{B}T})},
\end{equation}
with $T $ being the temperature of the resistor.
\noindent
If a second resistor is placed into the line, the above procedure can be applied independently to both resistors in order to construct the impedance matrix, and thus the exchanged heat power may be calculated as\cite{CavRevI}
\begin{equation} \label{eq:PowTransfer}
P_{k \rightarrow j}=\frac{R_{j}R_{k}}{\pi} \int_{-\infty}^{\infty} d\omega |Z^{-1}_{jk}(\omega)|^{2} \hbar \omega\left[f_{B}(-\omega , T_{j})-f_{B}(-\omega , T_{k})\right].
\end{equation}

\begin{figure} 
\includegraphics[width=0.5\textwidth]{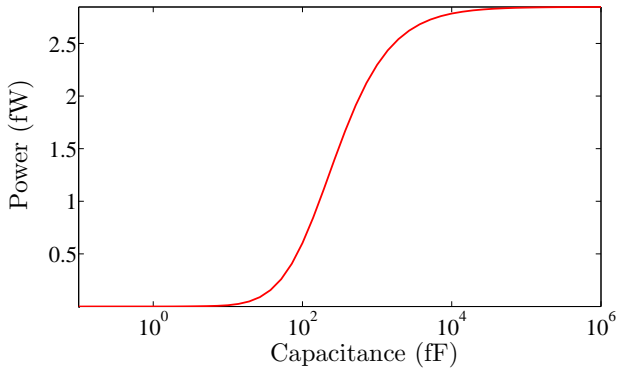}
\caption{Power transfer between two resistors, both of size $ 230 \textrm{ } \Omega$, situated at $ \frac{L}{10} $ and $ \frac{9L}{10} $ as a function of the magnitude of the capacitor placed at $ \frac{L}{2} $. The temperature of Resistor 1 is held at $ 100\textrm{ mK} $ while that of Resistor 2 is at $200 \textrm{ mK}$. The power saturates at $ 2.85 \textrm{ fW} $ in the limit $D \rightarrow \infty$. The parameters employed for the cavity are summarized in Table \ref{tbl:parameters}.}
\label{fig:POW1}
\end{figure}

The results for the power transfer between two resistors are shown in Fig.~\ref{fig:POW1} for the setup in which a capacitor in the center of the cavity is flanked by the two resistors at $ L/10 $ and $ 9L/10 $. The introduction of the capacitor reduces the power in comparison to the case of just one cavity. Note that with the resistors at the ends of the cavity, the coupling to the fundamental mode is weaker so that the imaginary part of the eigenfrequency no longer dominates for $ R=230 \textrm{} \Omega $.

\section{Tunable Heat Transport} \label{sec:SQUID}

In this section, we show that the power transfer between two resistors in a cavity may be tuned in situ by placing a SQUID into the line. The transition rates for the absorption and emission of photons between the cavity and the resistors, given in Eq.~(\ref{eq:TR}) are functions of the effective impedance of the center conductor. This suggests that by inserting SQUIDs into the line, we can change this effective impedance and therefore vary the coupling strength between the resistors in situ. 

Since we have only a few photons in the cavity we are comfortably in the low power regime. Following Ref.~\onlinecite{Bertet08},  we may then characterise a SQUID with critical current $ I_{c0} $, and loop-inductance $L_{\ell}$, as a linear inductor, provided $ L_{\ell} I_{c0}\ll \left(\Phi_{0}/2\pi\right)$. In this case the inductance of the SQUID is given by
\begin{equation}
L_{J}(\Phi)=\frac{\Phi_{0}}{2 \pi I_{c}(\Phi)},
\end{equation}
which depends on the flux through the SQUID via the effective critical current, $ I_{c}(\Phi)=2I_{c0}|\cos(\pi \Phi / \Phi_{0} )| $, with $ \Phi_{0}=\frac{h}{2e} $ being the flux quantum. In a real SQUID, even with negligible $ L_{\ell} $, there is always a slight assymmetry in the critical currents of the junctions $ I_{c1} $ and $ I_{c2} $. As a consequence of this assymmetry, $ I_{c} $ is bounded from below by $ |I_{c1}-I_{c2}| $ and is therefore never identically zero in practice\footnote{The lower bound given by the asymmetry can be circumvented using a so-called balanced SQUID with three Josephson junctions\cite{Kemppinen08}}. As $ L_{J} $ is constant for
a given $ \Phi $ we may, assuming small spatial dimensions for the SQUIDs, transform $Z(\omega) \rightarrow Z^{L}(\omega)$ simply by replacing $ L $ with $ L_{J}+L $ in the circuit shown in Fig. 5 at any point in the line in which we place a SQUID. This is valid if the capacitance of the SQUID, $C_{J}$, is small enough such that $\omega_{1} \ll 1/\sqrt{L_{J}C_{J}}$.

For simplicity we take the case in which we have only one SQUID in the line, which also distorts the mode profile. This effect is small for most values of $ \Phi $ but can be substantial at certain resonance points.
\begin{figure} 
\includegraphics[width=0.5\textwidth]{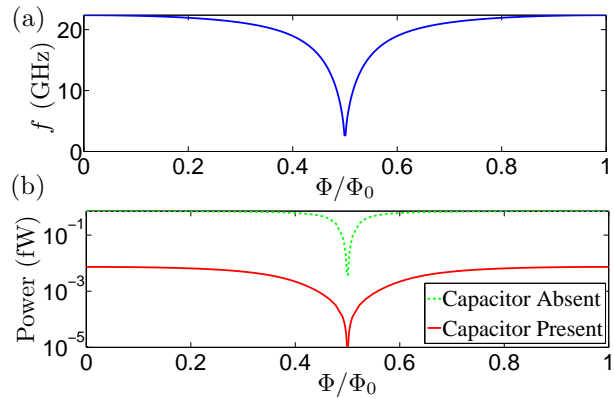}
\caption{(color online) Two resistors of size $ 230 \textrm{ } \Omega $ each are placed in the cavity at $ L/10 $ and $ 9L/10 $, between these resistors a SQUID of critical current $ I_{c0}=1.5 \textrm{ }\mu A $ is placed at $ L/4 $. (a) Fundamental eigenfrequency of the cavity as a function of the flux $ \Phi $ with a $10 \textrm{ fF}$ capacitor placed at $L/2$. (b) Power transfer between the resistors with (solid line) and without (dashed line) the $10 \textrm{ fF}$ capacitor. The results of both (a) and (b) are periodic in $ \frac{\Phi}{\Phi_{0}} $. The parameters employed for the cavity are summarized in Table \ref{tbl:parameters}.}
\label{fig:omphi}
\end{figure}
Again we may use Eq.~(\ref{eq:PowTransfer}) to calculate the photonic power transfer. In Fig.~\ref{fig:omphi}(a), the eigenfrequency of the cavity is shown as a function of the flux through the SQUID and Fig.~\ref{fig:omphi}(b) shows how the power can be modulated by the flux. We obtain peaks in the power exchange when on resonance. Although the maximum power may seem small, the power dissipated for a single photon in a very high quality factor cavity is of the order $ P_{1}\approx \frac{\hbar \omega^{2}}{Q}$. For illustration, a cavity with a quality factor of $1\times 10^{7}$ (see Ref. \onlinecite{Megrant12}), if we again take $f_{1}=1\times 10^{10}$, has $P_{1}\approx 0.1 \textrm{ aW} $ which is less than, or comparable with the powers observed in Fig.~\ref{fig:omphi}(b). An example of utilization would be to tune the cavity with the SQUID off resonance whilst we perform an experiment on the other half of the cavity. Afterwards, the SQUID cavity may be tuned back into resonance in order to reset the state of the high-Q cavity without the SQUID. For observation of tunable temperature changes in the resistor, the capacitor is not necessary and its removal increases the power by several orders of magnitude as shown in Fig.~\ref{fig:omphi}(b).
 \FloatBarrier
\section{Conclusions}
In summary, we have used a simple circuit model in order to demonstrate that the photonic power between two components in a cavity may be tuned on and off using a SQUID. Furthermore, cavity experiments often involve components which are coupled strongly to the cavity modes. These component can have a substantial impact on the mode profiles such that they no longer bear any relation to the sinusoidal profile of the bare cavity. We have also shown that SIN thermometry does not hinder the ability to observe single-photon heat conduction. 
\acknowledgments{We thank the Academy of Finland, the Emil Aaltonen Foundation, the Finnish National Doctoral Programme in Materials Physics (NGSMP) and the European Research Council under Grant Agreement No. 278117 (SINGLEOUT) for financial support.}
\bibliographystyle{apsrev}
\bibliography{Cav_Rev2}
\end{document}